\newif\ifmnras
\mnrastrue
\ifmnras
\documentclass[useAMS,usenatbib,usedcolumn]{mnras}
\else
\documentclass[twocolumn]{aastex62}
\fi

\usepackage{graphics,epsf}
\usepackage{amsmath}                
\usepackage{amsfonts}               
\usepackage{amssymb}                
\usepackage{epsfig}
\usepackage{epstopdf}
\usepackage{multirow}
\usepackage[para,online,flushleft]{threeparttable}

\usepackage{xcolor}
\definecolor{redak}{rgb}{0.9,0.15,0.05}

\def \kms{\, \mathrm{km} \, \mathrm{s}^{-1}}
\def \kev{\, \mathrm{keV}}

\def \s{\, \mathrm{s}}
\def \km{\, \mathrm{km}}

\def \erg{\, \mathrm{erg}}

\def \yr{\, \mathrm{yr}}

\def \rmModot{\, \mathrm{M_{\sun}}}
\def \rmRodot{\, \mathrm{R_{\sun}}}

\ifmnras
\title[Diversity of CEJSNe]{Diversity of common envelope jets supernovae and the fast transient AT2018cow}
\author[N. Soker, A. Grichener, A. Gilkis]{
Noam Soker$^{1,2}$\thanks{Contact e-mail: \href{soker@physics.technion.ac.il}{soker@physics.technion.ac.il}},
Aldana Grichener$^{1}$\thanks{Contact e-mail: \href{aldanag@campus.technion.ac.il}{aldanag@campus.technion.ac.il}}
Avishai Gilkis$^{3}$\thanks{Contact e-mail:  \href{agilkis@ast.cam.ac.uk}{agilkis@ast.cam.ac.uk}}
\\
$^{1}$ Department of Physics, Technion -- Israel Institute of Technology, Haifa 3200003, Israel\\
$^{2}$ Guangdong Technion Israel Institute of Technology, Shantou, Guangdong Province 515069, China\\ 
$^{3}$ Institute of Astronomy, University of Cambridge, Madingley Road, Cambridge, CB3 0HA, UK\\
}
\fi

\begin{document}

\ifmnras
\pagerange{\pageref{firstpage}--\pageref{lastpage}} \pubyear{2018}

\maketitle
\label{firstpage}
\else

\title{Diversity of common envelope jets supernovae and the fast transient AT2018cow}

\author[0000-0003-0375-8987]{Noam Soker}
\affil{Department of Physics, Technion -- Israel Institute of Technology, Haifa 3200003, Israel}
\affil{Guangdong Technion Israel Institute of Technology, Shantou, Guangdong Province 515069, China}
\correspondingauthor{N. Soker}
\email{soker@physics.technion.ac.il}

\author[0000-0002-7840-0181]{Aldana Grichener}
\affil{Department of Physics, Technion -- Israel Institute of Technology, Haifa 3200003, Israel}
\email{aldanag@campus.technion.ac.il}

\author[0000-0001-8949-5131]{Avishai Gilkis}
\affil{Institute of Astronomy, University of Cambridge, Madingley Road, Cambridge, CB3 0HA, UK}
\email{agilkis@ast.cam.ac.uk}

\shorttitle{Diversity of CEJSNe} 
\shortauthors{Soker, Grichener \& Gilkis}
\fi

\begin{abstract}
We propose a common-envelope jets supernova (CEJSN) scenario for the fast-rising blue optical transient AT2018cow. In a CEJSN a neutron star (NS) spirals-in inside the extended envelope of a massive giant star and enters the core. The NS accretes mass from the core through an accretion disc and launches jets. These jets explode the core and the envelope. In the specific \emph{polar CEJSN scenario} that we propose here the jets clear the polar regions of the giant star before the NS enters the core. The jets that the NS launches after it enters the core expand almost freely along the polar directions that contain a small amount of mass. This, we suggest, explains the fast rise to maximum and the fast ejecta observed at early times of the enigmatic transient AT2018cow. The slower later time ejecta is the more massive equatorial outflow. We roughly estimate the accretion phase onto the NS during the explosion phase to last for a time of $\approx 10^{3} \s$, during which the average mass accretion rate is $\approx 10^{-4} \rmModot \s^{-1}$. We outline the possible diversity of CEJSNe by listing five other scenarios in addition to the polar CEJSN scenario.
\end{abstract}

\ifmnras
\else
\fi

\section{INTRODUCTION}
\label{sec:intro}

The fast-rising blue optical transient AT2018cow was discovered on June 16, 2018 \citep{Prenticeetal2018} and has attracted attention since then (e.g., \citealt{Perleyetal2019, Hoetal2019, Kuinetal2019}). \cite{Marguttietal2019} conduct the most thorough study of this event, and compare it with the expectations of seven scenarios. They conclude that only three scenarios from those they consider are viable: an electron-capture supernova that formed a magnetar which further powered the supernova, a blue supergiant that collapsed to form a black hole (BH) which powered the supernova by accretion, and a core collapse supernova (CCSN) embedded in a dense torus of circumstellar matter (CSM). All scenarios require a highly aspherical explosion, and AT2018cow was most likely a bipolar explosion. The fast rise implies a very low mass in the ejecta which is observed with high velocities, of $\approx 0.1c$, at early days. Later the observed ejecta has lower velocities, of $\approx 4000 \, \mathrm{km} \, \mathrm{s}^{-1}$. In the above scenarios the polar ejecta is fast and of low mass, and the equatorial ejecta is slow and massive.   

\cite{LyutikovToonen2018} suggest that AT2018cow resulted from electron-capture collapse following a merger of two white dwarfs, one of them an ONeMg white dwarf.
   
\cite{Quataertetal2019} raise the possibility that the jittering jets explosion mechanism (e.g., \citealt{PapishSoker2011, GilkisSoker2015}) might explain the unique properties of AT2018cow. In their scenario the random angular momentum for the formation of the variable accretion disc that launches the jittering jets comes from the hydrogen-rich envelope. This is compatible with the property of the jittering jets mechanism that the explosion becomes more violent as farther out zones of the progenitor star supply the stochastic angular momentum to the accretion disc around the newly-born neutron star (NS) and then around the BH (\citealt{GilkisSoker2014}).

We suggest an alternative scenario, in which AT2018cow was a common envelope jets supernova (CEJSN) event, rather than a CCSN. The following properties bring us to propose this scenario: \linebreak 
\indent ($i$) The very rapid rise of AT2018cow indicates that the mass in the fast ejecta is very low, few$\times 0.1 M_\odot$ \citep{Marguttietal2019}, much lower than that expected from an explosion of a massive spherical star. We take it to imply that the star was highly distorted just before explosion. Namely, at explosion the polar directions were already of low mass. A binary companion far from the core of the exploding star cannot distort the inner stellar regions. For example, a binary companion outside the envelope of the exploding star that some studies (e.g., \citealt{McleySoker2014}) suggest to explain pre-explosion outbursts observed in many CCSNe (e.g., \citealt{Foleyetal2007, Ofeketal2013, SvirskiNakar2014, Moriya2015, Marguttietal2017, Yaronetal2017, Liuetal2018, Pastorelloetal2018}) cannot give the inferred clearance in the polar directions. It seems that the pre-explosion activity of the progenitor of AT2018cow was very different from the more commonly observed pre-explosion outburst activity. \linebreak 
\indent ($ii$) The requirement for a dense equatorial ejecta that is much slower than the polar ejecta \citep{Marguttietal2019} suggests either a rapid core rotation before explosion, or a common envelope evolution just before explosion. This also points to a strong binary interaction with the core just before explosion. \linebreak 
\indent ($iii$) The different characteristics of AT2018cow place it as a relatively rare event. We should consider then a rare scenario that is similar in some respects, but not identical, to other CCSNe.  

Overall, AT2018cow has some unique properties that we argue call for a rare binary interaction just before explosion. This raises the question of the coincidence between the binary interaction and the explosion that immediately followed.  We attribute both the pre-explosion and explosion mass ejections in such events to a neutron star that accretes mass from the envelope, and then from the core of a giant massive star.

We describe the CEJSN scenario in section \ref{sec:scenario}. In section \ref{sec:core} we show that in many cases the NS is likely to enter the core during the common envelope evolution phase, and in section \ref{sec:accretion} we study the accretion rate onto the NS as it spirals-in inside the core, and from that we estimate the available energy and time scales. In section \ref{sec:Charectaristic} elaborate on the characteristics of the polar CEJSN scenario for AT2018cow. In section \ref{sec:summary} we summarise our results that call for a greater attention to the CEJSN scenario when studying peculiar supernovae and supernova impostors.

\section{THE PROPOSED CEJSN SCENARIO}
\label{sec:scenario}

\subsection{Properties of the CEJSN scenario}
\label{subsec:basic}

The CEJSN scenario is based on the possibility of a NS that orbits inside a massive giant star to accrete mass at very high rates thanks to cooling by neutrinos when the mass accretion rate is $\dot M_{\rm acc} \ga 10^{-3} \rmModot \yr^{-1}$ \citep{HouckChevalier1991, Chevalier1993, Chevalier2012}, and the likely accretion through a disc \citep{ArmitageLivio2000, Papishetal2015, SokerGilkis2018}. The accretion disc might launch jets, and the interaction of the jets with the envelope and the circumstellar matter (CSM) converts kinetic energy to thermal energy. The outcome might be an explosion mimicking CCSNe, but with axisymmetric, rather than spherical, CSM \citep{Chevalier2012}. We note that a BH companion can have similar effects to that of a NS, but here we focus mainly on NS companions.

\cite{SokerGilkis2018} propose that the enigmatic supernova iPTF14hls \citep{Arcavietal2017} was a CEJSN event. In this scenario the NS first accretes mass from the tenuous envelope, leading to a dense axisymmetric CSM, and then interacting with the core, powering the explosion by launching jets. In a CEJSN event the NS destroys the core. If the NS accretes mass from the tenuous envelope to power a bright transient event, but it does not spiral-in all the way to the core, the event is termed a CEJSN impostor \citep{Gilkisetal2019}.

In addition to accounting for asymmetrical CSM \citep{Chevalier2012, SokerGilkis2018}, \cite{Papishetal2015} and \cite{GrichenerSoker2019} propose that when the giant is at a late evolutionary stage and has a CO core, the CEJSN might be a site of r-process nucleosynthesis of heavy elements. The nuclear reactions take place inside the jets. In the present study we consider a NS that enters the core at an earlier stage, when the core density is lower and the accretion rate is relatively low such that the material in the jets is not neutron-rich. We do not expect r-process nucleosynthesis to take place. 
   
\begin{figure*}
\begin{center}
\includegraphics[width=1\textwidth]{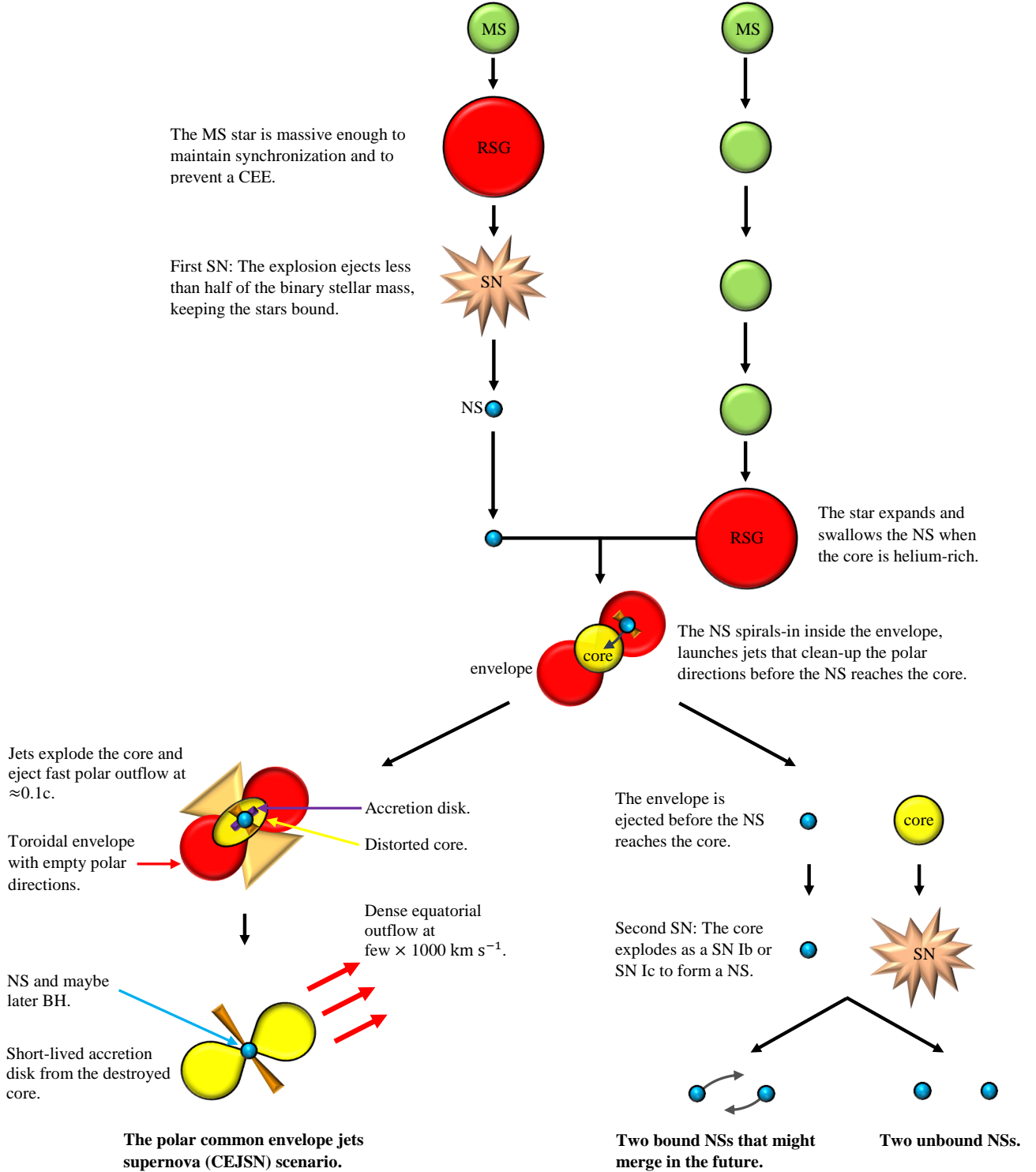}
\vspace*{-2.2cm}
\caption{A schematic drawing of the polar CEJSN scenario we propose for AT2018cow. MS denotes a main sequence star, and RSG stands for red supergiant star. The initially less massive star swallows the NS during its early expansion phase, when it has a helium core. The NS might end outside the helium core, followed by a Type Ib or Type Ic CCSN that forms a NS binary system, bound or unbound (right-hand side). We are instead interested in the case where the NS enters the core and explodes the star (left-hand side).}
\label{fig:CEJSN}
\end{center}
\end{figure*}

In our proposed scenario the jets that the NS launches as it approaches the core empty the stellar mass from the two opposite polar directions. We term this scenario the \textit{polar CEJSN scenario}. As the NS enters the core it accretes mass at a much higher rate, and hence it launches very energetic jets. These jets power the explosion and explain the fast low-mass ejecta, with a velocity of $\approx 0.1 c$, that is observed in the first days. Later the destroyed core forms an accretion disc around the NS. This leads to further accretion and the launching of jets that power AT2018cow at later times. We schematically draw our proposed scenario in Fig. \ref{fig:CEJSN}. 

There are uncertainties in the accretion rate and in the formation of an accretion disc as the NS spirals-in inside the envelope of giant progenitor. Studies of compact objects spiraling-in inside stellar envelopes, and in particular hydrodynamical simulations, obtain different values for the mass accretion rate, and come to different conclusions on the question of whether an accretion disc forms or not around the compact object (e.g., \citealt{RasioShapiro1991, Fryeretal1996, Lombardietal2006, RickerTaam2008, MacLeod2015, MacLeod2017}), 

\cite{MacLeodRamirezRuiz2015b}, for example, argue that the steep envelope density gradient reduces the accretion rate. On the other hand, \cite{Staffetal2016}, who include envelope rotation in their 3D simulations, find that the mass flowing from the giant to the compact object before the latter enters the envelope forms an accretion disc. Once an accretion disc exists it is likely to launch jets, which remove angular momentum and sustain the disc. \cite{Chamandyetal2018} do not include envelope rotation in their 3D simulations, but use a sub-grid scheme to remove mass and energy from the vicinity of the accreting body (as we expect jets to do; e.g., \citealt{Shiberetal2016}), find the formation of an accretion disc inside the common envelope.

There are also hints that even main sequence stars might launch jets inside a common envelope.  \cite{BlackmanLucchini2014} explain the large momenta in some bipolar planetary nebulae as energetic jets that a main sequence companion launches inside a common envelope. In the present study the companion is a NS that is much smaller than a main sequence star, hence the accreted gas is much more likely to form an accretion disc. 
 
Overall, we consider the formation of an accretion disc around a NS spiraling-in inside the envelope of a giant star to be extremely likely, and so is the launching of jets by this accretion disc.

\subsection{The importance of the envelope mass}
\label{subsec:envelopemass}

In the proposed CEJSN scenario for iPTF14hls \cite{SokerGilkis2018} assume a very massive envelope, and the NS cannot deposit sufficient angular momentum to the envelope to transform it into a torus. Also, as the envelope is very massive, the jets launched from the NS do not penetrate the envelope. Therefore, the jets do not empty the polar regions.

In our proposed model for AT2018cow here, on the other hand, the mass of the envelope is smaller and the NS can eject a large fraction of the envelope in the equatorial plane to form a torus. The jets can then penetrate along the polar directions of the envelope and empty those regions for the later ejecta to flow almost freely along the polar directions. This accounts for the fast early ejecta in AT2018cow. 

\subsection{The key role of the accretion rate}
\label{subsec:variationrate}

A key parameter of the CEJSN is the variation of the mass accretion rate onto the NS with time, and the interaction of the NS, and the jets it launches, with the envelope and core. Another parameter is the accretion rate onto the NS when it accretes material from the core. There are many possible combinations, and we discuss only several. 

When the accretion rate is very high the material in the accretion disc is sufficiently dense for electron degenerate energy to be above the threshold for $p+e \rightarrow n+\nu_e$. As a result of that the jets are highly neutron-rich, and r-process nucleosynthesis might take place inside the jets \citep{Papishetal2015, GrichenerSoker2019}. This is termed the CEJSN r-process scenario. 

In cases where the NS enters the envelope but not the core then there is an outburst, but not a terminal explosion. Such cases take place when the NS manages to eject the entire envelope before it spirals-in into the core, or when it enters the envelope on an eccentric orbit and exits the envelope. This energetic outburst leads to an intermediate luminous optical transient (ILOT) that is termed a CEJSN impostor \citep{Gilkisetal2019}. Later the core explodes and leaves behind another NS, or BH, either bound or not to the first NS.

We also define the \textit{standard CEJSN} scenario, in which the duration and energy of the explosion are similar to typical CCSNe. The light curve might be different and depend on viewing angles because of a bipolar explosion. In this case both the CSM and the envelope at explosion are bipolar, but the polar directions are not empty, hence it is not a polar CEJSN.  

In Table \ref{Table:Scenarios} we summarise the different CEJSN scenarios. For the prolonged CEJSN, which has the attributes pertinent to iPTF14hls, we require that a large mass stays bound when the NS enters the core. For a CO core that has a small radius the liberated orbital energy is very large, and likely to unbind most of the envelope, so for the prolonged CEJSN a helium-rich core is necessary. Numerical simulations will find the precise boundaries between the different scenarios. In the table we give our crude preliminary estimates for some parameters.    
\begin{table*}
\caption{CEJSN Scenarios}
\centering
\begin{threeparttable}
\begin{tabular}{| l | c | c | c | c | c |}
\hline
Name     & Pre-explosion  & Envelope      & Pre-explosion & Post-explosion    & Outcome     \\ 
         & envelope       & at explosion  & core          & core              &             \\
\hline
Polar    & Low mass;      & Unbound bipolar   & Helium-rich;  & Mostly unbound    & $E_{\rm exp} \approx 10 E_{\rm bind}$;\\ 
CEJSN    & highly oblate; & CSM; dense torus. & large.        & ejecta; colliding & Fast low-mass polar ejects; \\ 
         & empty poles.   &                   &               & with CSM          & NS/BH; \textbf{AT2018cow}.     \\ 
          \hline   
Standard&$\approx 12$--$30 \, \mathrm{M}_\odot$;& bipolar   & Helium or CO; & Little mass is    &$E_{\rm exp} \approx 10 E_{\rm bind}$; energetic  \\ 
CEJSN    & oblate        & dense CSM; oblate  & Massive.      & bound; a typical & ejecta-CSM collision; mimicking \\ 
         &               & at explosion.   &                  & CCSN duration    & a CCSN of type II; NS/BH. \\ 
   \hline   
Stripped- &$\la 1\rmModot$;& bipolar low-  & CO-core; & Little mass is    &$E_{\rm exp} \approx 10 E_{\rm bind}$; energetic  \\ 
envelope & oblate        & density CSM; oblate  & Massive.      & bound; a typical & ejecta-CSM collision; mimicking \\ 
CEJSN    &               & Helium-rich.          &                  & CCSN duration    & a CCSN of type Ib; NS. \\ 
   \hline   
   
Prolonged&Very-high mass;& Mildly bipolar    & Helium-rich;  & Partly bound,  & $E_{\rm exp} \gg E_{\rm bind}$; Very \\ 
CEJSN    &weakly oblate. & dense CSM.        & very massive. & causing prolonged & energetic ejecta-CSM collision; \\ 
         &               &                   &               & accretion         & NS/BH; \textbf{iPTF14hls}.\\ 
   \hline   
 CEJSN   &Extended;      & Bipolar CSM; low  &Evolved, e.g., & Neutron-rich jets & $E_{\rm exp} \gg E_{\rm bind}$; \\ 
r-process&ejected by     & density along     & CO-core and   & + ejecta as in    & r-process nucleosynthesis\\ 
         &jets.          & polar directions. & beyond        & above cases       & inside neutron-rich jets. \\ 
   \hline   
   
CEJSN    &Extended       & Intact;           & Any type      & Unchanged as there& ILOT that might repeat, e.g.,  \\ 
impostor &               & no explosion,     &               & is no explosion.  & similar to  \textbf{SN~2009ip}; \\ 
         &               & only outburst.    &               &                   & (later leaves two NSs or BHs). \\ 
   \hline   
\end{tabular}
\footnotesize
\begin{tablenotes}
$E_{\rm exp}$ is the explosion energy, including kinetic energy and radiation, and $E_{\rm bind}$ is the binding energy of the star. NS/BH means a NS or BH remnant. 
\end{tablenotes}
\end{threeparttable}
\label{Table:Scenarios}
\end{table*}

\section{Entering the core}
\label{sec:core}
 
We follow with the Modules for Experiments in Stellar Astrophysics (MESA; version 9575; \citealt{Paxtonetal2011,Paxtonetal2013,Paxtonetal2015,Paxtonetal2018}) \texttt{star} module the evolution of three stellar models to show that under some circumstances a NS companion can spiral-in all the way to the core. For the polar CEJSN scenario, where the NS manages to almost empty the polar directions, we require a stellar mass which is low compared to typical CCSN progenitors. For lower stellar masses the NS is more likely to eject the envelope before reaching the core, and then there will be no energetic supernova-like transient event. We therefore examined stellar models with initial masses $M_{\rm ZAMS}=8$, $11$, and $15 \rmModot$, all with an initial metallicity of $Z=0.014$.

We use the $\alpha$-prescription for common envelope evolution to calculate the final orbital separation of the core-NS system. This prescription takes that a fraction $\alpha$ of the energy that the spiralling-in NS liberates is channelled to remove the envelope that has a binding energy of $E_{\rm bind, e}$. The final orbital separation is  
\begin{equation}
a_{\rm f}
\simeq \frac{\alpha GM_{\rm core}M_{\rm NS}}{2E_{\rm bind, e}} ,
\label{eq:afinal}
\end{equation}
where $M_{\rm core}$ is the mass of the core and $M_{\rm NS}$ is the mass of the NS that we take here to be $M_{\rm NS}=1.4 \rmModot$. If the final orbital separation is smaller than the radius of the core, $a_f < R_{\rm core}$, we consider the NS to spiral-in into the core and then destroy the core.  
 
We study three values of $\alpha$ according to the traditional prescription where $\alpha \le 1$. We also run one `toy model' to take into account the effect that jets might have. As the NS spirals-in within the envelope it is likely to accrete mass and launch jets. These jets can play a major role in removing envelope mass (see section \ref{subsec:basic}). To mimic this possibility we present the results for one case with $\alpha=3$. Namely, the jets deposit energy which is twice that of the orbital energy into removing the envelope.  
 
\begin{figure}
\begin{center}
\includegraphics[width=0.49\textwidth]{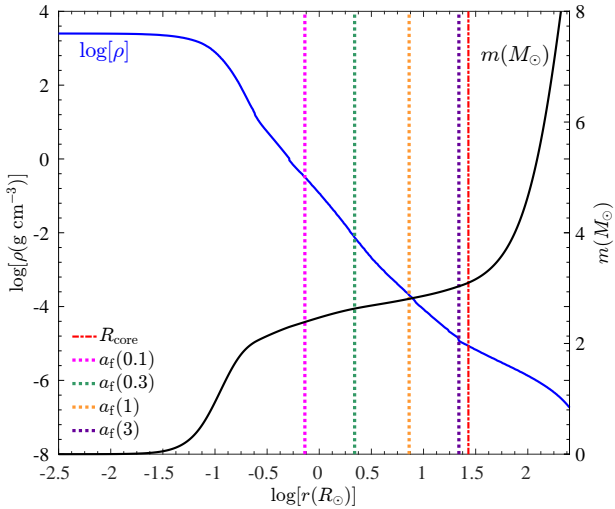}
\vspace*{-2.5cm}
\caption{Mass (black) and density (blue) profiles of our $M_{\rm ZAMS}=8 \rmModot$ stellar model assuming the NS enters the envelope of the giant star when the radius of the giant is $200 \rmRodot$. The radius of the the helium core is marked by the red dash-dotted line. The orbital separation between the NS and the centre of the core at the end of the common envelope phase for $\alpha=0.1$, $\alpha=0.3$, $\alpha=1$,and  $\alpha=3$ are denoted by the magenta, green, orange, and purple dotted lines, respectively. Results for times when the giant is smaller also end with a NS-core merger. The binding energy of the envelope and core in this model are $E_{\rm bind, e}=6.9 \times 10^{47} \erg$ and $E_{\rm bind, c}= 1.4 \times 10^{49}\erg$, respectively.}
\label{fig:MassDensity8M}
\end{center}
\end{figure}

In Fig. \ref{fig:MassDensity8M} we present the model of the $M_{\rm ZAMS}=8 \rmModot$ star when it reaches a radius of $R=200 \rmRodot$. We also present the final orbital separation for four different values of the common envelope $\alpha$ parameter. In all four cases the NS enters the core. We examined cases when the giant swallows the NS at earlier times, when its radius is smaller, down to $50 \rmRodot$, and found no significant differences. In all cases when the giant swallows the NS the NS spirals-in all the way into the core.

\begin{figure}
\begin{center}
\includegraphics[width=0.49\textwidth]{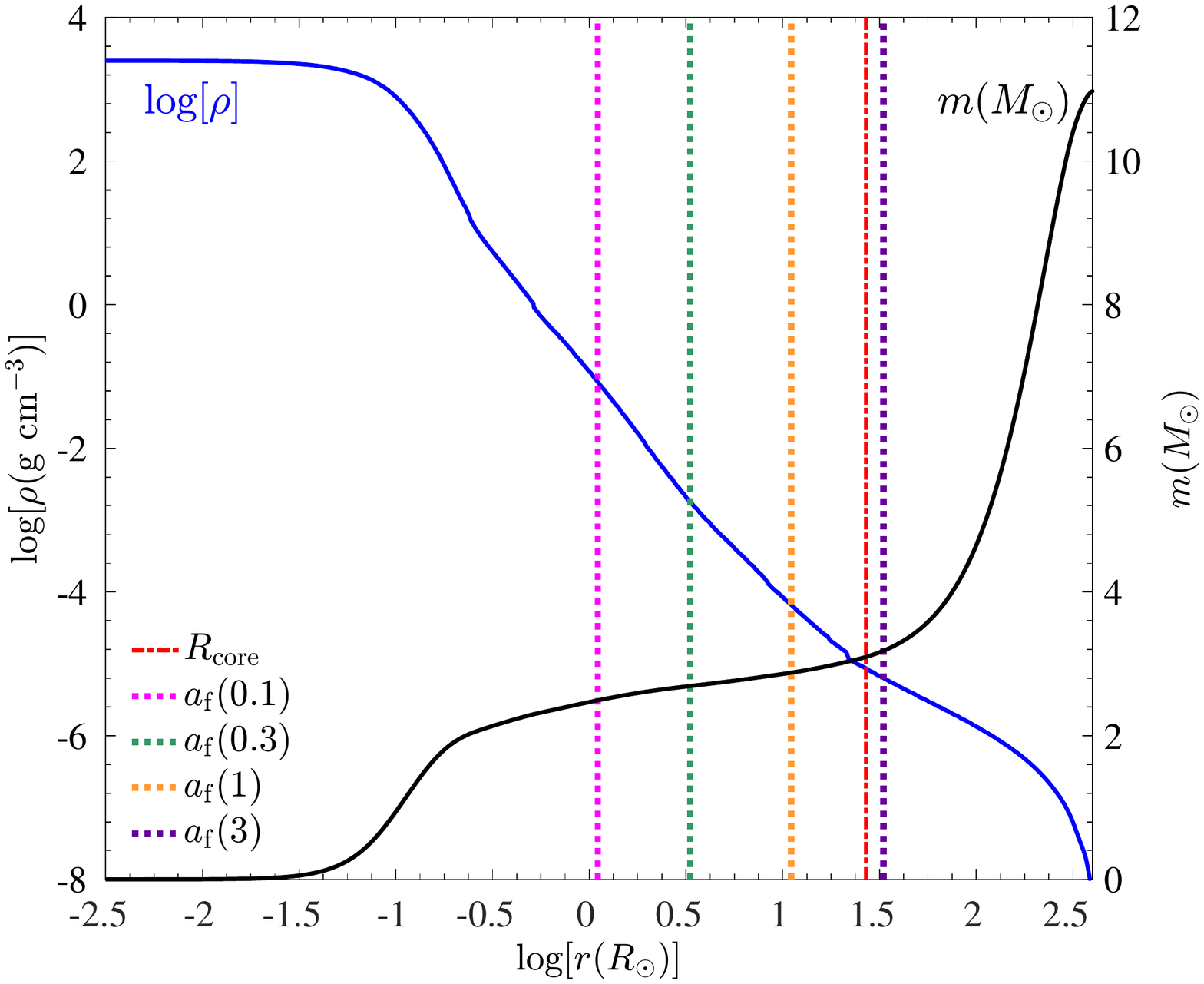}
\vspace*{-2.5cm}
\caption{Like Fig. \ref{fig:MassDensity8M} but for the $M_{\rm ZAMS}=11 \rmModot$ stellar model assuming the NS enters the envelope of the giant star when the radius of the giant is $400 \rmRodot$. The binding energy of the envelope and core in this model are $E_{\rm bind, e}=6.8 \times 10^{47} \erg$ and $E_{\rm bind, c}= 3.2 \times 10^{49}\erg$, respectively.}
\label{fig:MassDensity11M400R}
\end{center}
\end{figure}

\begin{figure}
\begin{center}
\includegraphics[width=0.49\textwidth]{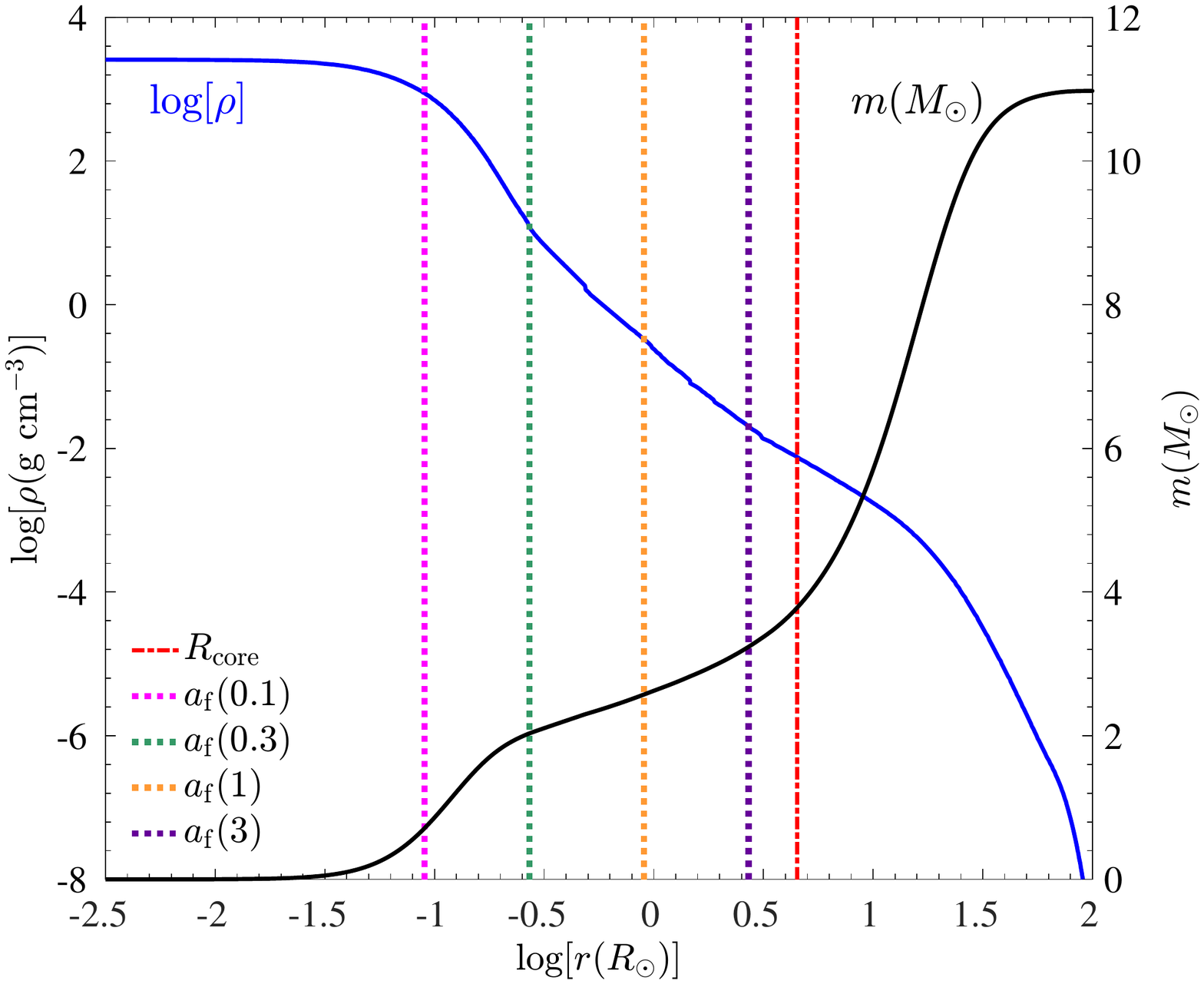}
\vspace*{-2.5cm}
\caption{Like Fig. \ref{fig:MassDensity8M} but for the $M_{\rm ZAMS}=11 \rmModot$ stellar model assuming the NS enters the envelope of the giant star when the radius of the giant is $100 \rmRodot$. The binding energy of the envelope and core in this model are $E_{\rm bind, e}=9.4 \times 10^{48} \erg$ and $E_{\rm bind, c}= 3.3 \times 10^{49}\erg$, respectively. In this case the NS falls into the core for $\alpha=3$.}
\label{fig:MassDensity11M100R}
\end{center}
\end{figure}

In Fig. \ref{fig:MassDensity11M400R} we present the stellar model with $M_{\rm ZAMS}=11 \rmModot$ when it reaches a radius of $R=400 \rmRodot$. We also present the final orbital separation for four different values of the $\alpha$ parameter. In the case of $\alpha=3$, namely where we assume that the jets deposit twice as much energy as the orbital energy, the NS does not enter the core. We examined cases when the giant swallows the NS at earlier times, when its radius is smaller. For a stellar radius smaller than about $100 \rmRodot$ the NS spirals-in into the core for $\alpha=3$, as we present in fig. \ref{fig:MassDensity11M100R}. The role of the jets allows for a large parameter space for which the NS survives, and the system later forms a double NS system (Fig. \ref{fig:CEJSN}). 
  
\begin{figure}
\begin{center}
\includegraphics[width=0.49\textwidth]{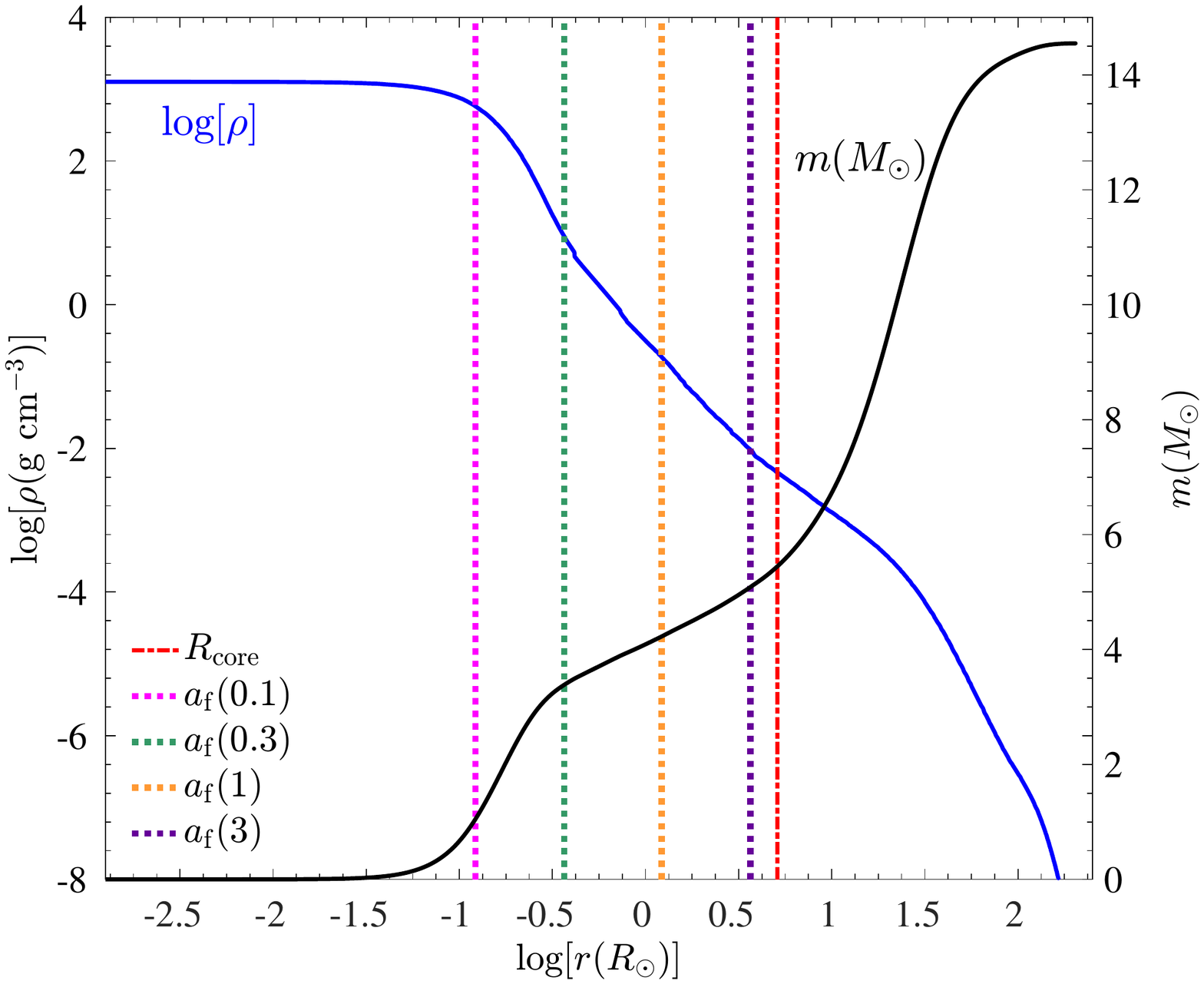}
\vspace*{-2.5cm}
\caption{Like Fig. \ref{fig:MassDensity8M} but for the $M_{\rm ZAMS}=15 \rmModot$ stellar model assuming the NS enters the envelope of the giant star when the radius of the giant is $200 \rmRodot$. The binding energy of the envelope and core in this model are $E_{\rm bind, e}=9.9 \times 10^{48} \erg$ and $E_{\rm bind, c}= 6.3 \times 10^{49} \erg$, respectively. }
\label{fig:MassDensity15M}
\end{center}
\end{figure}

In Fig. \ref{fig:MassDensity15M} we present the model and final orbital separations for the $M_{\rm ZAMS}=15 \rmModot$ stellar model. In this case when the NS enters the envelope at later stages, when the envelope is larger, it might not reach the core even for $\alpha<1$. 

Our conclusion from this section is that in many cases the NS enters the core if the star which swallows the NS has the appropriate characteristics for the polar CEJSN scenario. The core is helium-rich and has a mass of $M_{\rm core} \simeq 2$--$4 \rmModot$, and the radius of the core varies from $R_{\rm core}\simeq 3 \rmRodot$, for massive stars, up to much larger radii, of $R_{\rm core}\simeq 30 \rmRodot$, for lower mass stars. In the next section we examine the mass accretion rate onto the NS.

\section{THE ACCRETION PROCESS}
\label{sec:accretion}

We follow the study of \cite{GrichenerSoker2019} (where more details can be found) and estimate the accretion rate onto the NS by using the Bondi-Hoyle-Lyttleton mass accretion rate from the undisturbed core. This overestimates the accretion rate in the outer regions of the core because the jets from the NS remove mass from the NS vicinity and eject mass from the core. In regions of the core where the mass inner to the orbital separation is not much larger than the NS mass, the NS destroys the core.

\begin{figure}
\begin{center}
\includegraphics[width=0.49\textwidth]{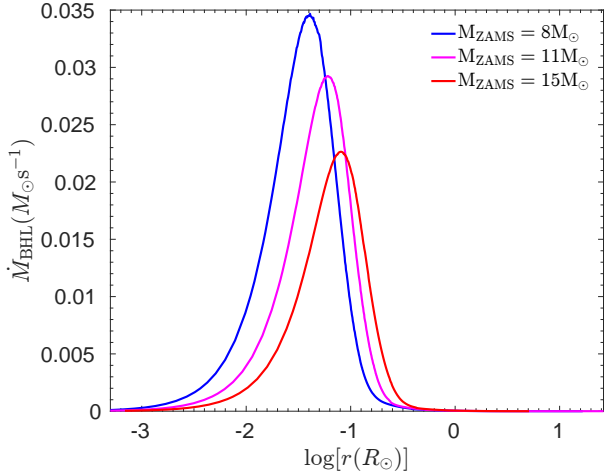}
\vspace*{-2.5cm}
\caption{The Bondi-Hoyle-Lyttleton  mass accretion rate onto the NS as it spirals-in inside the core of our $M_{\rm ZAMS}=8 \rmModot$, $M_{\rm ZAMS}=11 \rmModot$ and $M_{\rm ZAMS}=15 \rmModot$ stellar models, as presented in Figs. \ref{fig:MassDensity8M}--\ref{fig:MassDensity15M}. For our $M_{\rm ZAMS}=11 \rmModot$ we show the accretion rate in the case where the NS enters the envelope at the end of the first expansion, yet the results for smaller radii are not much different.}
\label{fig:Accretion}
\end{center}
\end{figure}

During the entire process the NS accretes a small fraction of the core mass, and ejects most of the core mass. One should take into account these approximations when analysing the accretion rate we present below. In Figs. \ref{fig:Accretion} we present the accretion rates under our assumptions for the three stellar models as presented in Figs. \ref{fig:MassDensity8M}-\ref{fig:MassDensity15M}. 

The orbital period of the NS-core binary system at a separation of $a \simeq 0.1 \rmRodot$ is about $P\simeq 200 \s$. At an accretion rate of $0.02 \rmModot \s^{-1}$ (Fig. \ref{fig:Accretion}) the NS would accrete the core in less than one orbital period. We show that in principle the NS can accrete at a very high rate, but the accretion rate will be lower than the values presented in Fig. \ref{fig:Accretion} due to the negative feedback nature of the interaction. Namely, the jets remove mass from the NS vicinity and the envelope, and by that reduce the accretion rate. Later the NS destroys the core.

More reasonable values for the interaction will result in an evolution that lasts for tens of minutes before the core is destroyed, and a total accreted mass that leads to the destruction of the core. Most of the accretion energy ends in neutrinos that escape, and only a small fraction goes to the kinetic energy of the jets. We can estimate the total mass the NS accrete as follows. We expect the total energy of the jets to be several times the binding energy of the core, $E_{\rm exp} = q E_{\rm bind, c}$ with $q \approx 10$, and where the kinetic energy of the jets is actually the explosion energy of the CEJSN. In super-energetic supernovae the ratio can be even $q \approx 100$. This is an important point, as requiring the jets to unbind the core until accretion ceases makes the CEJSN having the same order of magnitude explosion energy as regular CCSNe. In cases of jets that expand freely and do not unbind mass in the equatorial plane the explosion energy can be larger even. The large helium-rich core, which is more dispersed relative to a CO core, allows a lower explosion energy of the CEJSN in that case. So the freely-expanding jets in the polar CEJSN scenario make the value of $q$ larger. Overall, the explosion energy is $E_{\rm exp} \approx 10^{51} \erg$. 

Finally, after the NS enters the core the ejecta has to overcome the gravitational potential of the NS, in addition to the gravity of the core. This increases the binding energy of the ejecta, but we cannot calculate it without full 3D hydrodynamical simulations. 

If the NS launches in the jets a fraction of $\eta \simeq 0.1$ of the mass it accretes at a velocity of $v_j \simeq 10^5 \kms$, then we require the total mass the NS accretes to be 
\begin{eqnarray}
\begin{aligned}
M_{\rm acc}  \simeq & 
0.1
\left( \frac{q}{10} \right)
\left( \frac{\eta}{0.1} \right)^{-1} 
\\ & \times 
\left( \frac{E_{\rm bind, c}}{10^{50} \erg} \right) 
\left( \frac{v_j }{10^5 \km \s^{-1}} \right)^{-2} \mathrm{M}_\odot
\end{aligned}
\label{eq:Macc}
\end{eqnarray}
In super-energetic supernovae, when $q \ga 100$, the NS might increase its mass enough to collapse to a BH.

For an accretion phase that lasts for several orbital periods, $\tau_{\rm acc,NS} \approx {\rm several} \times P \approx 10^3 \, \mathrm{s}$, the accretion rate is  $M_{\rm acc,NS} \approx 10^{-4} \rmModot \s^{-1}$. The implication of Fig. \ref{fig:Accretion} is simply that the density in the core allows this high accretion rate. Although the accretion phase is much longer than that in regular CCSNe, the time of $\tau_{\rm acc,NS} \approx 10^3 \s$ is still several orders of magnitude shorter than the shock propagation time in the envelope of the star, and hence has no direct observational consequences on the light curve at explosion. Yet the distribution of newly synthesised elements in the ejecta might be different. To explore the distribution of the synthesised elements we need comprehensive 3D hydrodynamical simulations.  

\section{AT2018cow and the characteristics of the polar CEJSN scenario}
\label{sec:Charectaristic}

We now elaborate on the characteristics of our proposed polar CEJSN scenario for AT2018cow, and note that from all previous scenarios and analyses of AT2018cow, we are closest to the general picture described by  \cite{Marguttietal2019}. Like \cite{Marguttietal2019} we also have a bipolar flow, here both the explosion and the CSM are bipolar, and like them we have a central engine. In our case the central engine is the NS that entered the envelope of the giant star, rather than the newly born NS or BH. We also differ by attributing the explosion to jets that the NS launches, rather than to a neutrino driven supernova. Therefore, for most properties of AT2018cow we account in similar ways to those suggested by \cite{Marguttietal2019}. For example, like \cite{Marguttietal2019}, we attribute the X-ray variability to changes in the power of the central engine. In our case it is the launching of jets that is accompanied by X-ray emission (see below).
 
\cite{Marguttietal2019} require the flow to be highly bipolar, e.g., a fast low-density polar outflow and a slower denser equatorial outflow. While \cite{Marguttietal2019} do not specify the reason for this bipolar outflow in their model, in our polar CEJSN scenario the bipolar outflow is a natural outcome of the binary interaction before explosion and the jets that the NS launches before, during, and after the explosion.  
 
 We list some of the characteristics of our scenario, but emphasise that there are still large uncertainties for estimating these.

\textit{Occurrence rate.} \cite{Chevalier2012} estimated the rate of events in which a NS enters the envelope of a giant from the results of \cite{Podsiadlowski1995} to be about $1 \%$ of the rate of CCSNe.  \cite{Chevalier2012} further concluded that in most cases the NS enters the core, and the system does not leave behind a binary NS or NS-BH system. For a CCSN rate in the local universe of $\approx 1.5 \times 10^{5} {\rm Gpc}^{-3} \yr^{-1}$ (e.g., \citealt{Mattilaetal2012}) and a NS-NS merger rate of $\approx 1.5 \times 10^{3} {\rm Gpc}^{-3} \yr^{-1}$ (e.g., see discussion by \citealt{GiacobboMapelli2018}), there are $\approx 10$ NS-NS merger events per 1000 CCSN events. The occurrence rate of CEJSN events might be $\approx 10$--$30$ events per 1000 CCSN events. From these, we roughly estimate that the occurrence rate of each of the CEJSN channels that we list in Table \ref{Table:Scenarios} should be about several events per 1000 CCSNe. In other words, we crudely estimate that $\approx 2$--$5$ polar CEJSN events, i.e., AT2018cow-like events, take place for 1000 CCSN events. The next step will be to conduct a population synthesis study of the entire CEJSN class.  
         
\textit{Host galaxies.} AT2018cow is in the galaxy CGCG~137-068 which is a star-forming galaxy \citep{Prenticeetal2018}. \cite{Perleyetal2019} conclude that this galaxy is a star-forming dwarf spiral similar to the Large Magellanic Cloud. We can account for that as follows. In their population synthesis study \cite{Mapellietal2018} find that NS-NS mergers in the local Universe tend to occur in massive galaxies and shortly after star formation. Most CEJSN events occurs in the same type of stellar binary systems that lead to NS-NS binary systems, and so are expected to take place in star-forming galaxies. As well, a small fraction of CEJSN events might come from systems that have a more massive core that might end as a BH if there is no core-NS merger. Hence, some CEJSN events might take place in low-mass galaxies. In short, CEJSNe might take place in all types of star-forming galaxies. Future population synthesis will determine the expectation rate in different types of galaxies. 
   
\textit{Hydrogen spectrum.} The removal of envelope mass with jets implies the formation of a bipolar outflow, e.g., as observed in many planetary nebulae. In such a bipolar ejection the polar outflow is fast and of a relatively low density, while the equatorial outflow is slower and denser. The polar outflow is responsible for the absorption lines that indicate an outflow with a velocity of $\approx 0.13c$ (e.g., \citealt{Hoetal2019}), and the slower, about several$\times 1000 \km \s^{-1}$, equatorial outflow is responsible in our scenario for the broad hydrogen emission lines that appear later as reported by, e.g., \cite{Perleyetal2019}. Our explanation is essentially very similar to that of \cite{Marguttietal2019}.
    
\textit{Early non-thermal X-ray features.} There is a non-thermal hard X-ray emission ($>15 \kev$) at early times, i.e., about the first two weeks \citep{Marguttietal2019}. \cite{Marguttietal2019} present a model for the X-ray emission. As we indicated above, our flow geometry with a central engine is similar to that of \cite{Marguttietal2019}. For example, we also have disc-reprocessed emission from an equatorial disc. There is also emission from the jets. \cite{Kuinetal2019} attribute the non-thermal X-ray emission to jets in a tidal disruption event. We expect the jets in our polar CEJSN scenario to have non-thermal X-ray emission. Jets that the NS launches in X-ray binaries have non-thermal X-ray emission, with contributions from the base of the jets and from the accretion disc corona (e.g., \citealt{Migliarietal2010}). The variability in the accretion rate on to the accretion disc and from the accretion disc on to the NS and its characteristics, like the partition of energy between kinetic energy of the jets and between radiation, account for the engine variability. 

\textit{Late infrared (IR) increase.} There is an IR excess \citep{Perleyetal2019}, attributed by \cite{Kuinetal2019} to free-free emission from an extended optically-thin expanding atmosphere. However, \cite{Marguttietal2019} argue that this process cannot explain the temporal variability of the near-IR reported by \cite{Perleyetal2019}. The variability is over a time scale of days and manifests itself as several bumps in the light curve. We speculate that the IR excess is due to free-free emission in an extended expanding atmosphere, and the time variability is due to the interaction of variable jets with the main atmosphere and/or a highly non-smooth CSM into which the ejecta runs. This is a subject of a future study.

\textit{Optical and X-ray afterglows of similar luminosity.}  Several processes determine the X-ray and optical light curves in the proposed model. The ratio of X-ray to optical emission can vary a lot in different CEJSN. In the polar CEJSN scenario that we propose to explain AT2018cow the highly bipolar flow determines the energy partition between X-ray and longer wavelengths emission, similarly to the flow structure that \cite{Marguttietal2019} describe. We expect to have both an interaction of the ejecta with a bipolar CSM (with denser equatorial gas) and a central X-ray source as we described above. The transition from X-ray to UV/visible/IR emission is accounted for by the receding of the photosphere from the fast polar ejecta to the slower equatorial ejecta, as \cite{Marguttietal2019} depict in their Figure 17. 

\textit{The relation of AT2018cow to other supernovae.} We call attention to a paradigm in which most (or even all) CCSNe are driven by jets (e.g., \citealt{Soker2019RAA} for a recent paper) as an alternative to the questionable delayed neutrino-driven mechanism. Clearer cases that require energetic jets are super-energetic CCSNe \citep{Gilkisetal2016} and CCSNe where a magnetar supplies extra energy after explosion \citep{SokerGilkis2017}. In that sense the explosion by jets might not be an unusual property, but rather the unusual property is that the NS that launches the jets is not a newly-born NS from the core collapse, but an old NS that entered the envelope of the giant star. This process forms the class of unusual CEJSNe that we summarised in Table \ref{Table:Scenarios}. Different CEJSN channels might account for other unusual supernovae, such as iPTF14hls \citep{SokerGilkis2018}. We emphasise that the paradigm of jet-driven supernovae can account for both regular CCSNe and to many peculiar (unusual) exploding massive stars. 

\section{SUMMARY}
\label{sec:summary}

We examined the possibility that one class of the CEJSN scenario might account for the enigmatic transient (supernova) AT2018cow. Our more extended goal is to advance the exploration of the CEJSN scenario and its likely diversity, and the prospect that it might account for some rare peculiar explosions and transient events. In studying AT2018cow we were motivated by the fast ejecta at early times, the rapid rise to maximum luminosity, the slower ejecta at later times, and the overall enigmatic nature of this supernova (section \ref{sec:intro}) which makes it a rare-class event. We schematically depict the \textit{polar CEJSN scenario} that we propose here in Fig. \ref{fig:CEJSN}.

In the polar CEJSN scenario the giant star is not too massive as to allow the NS to clear the polar directions before it enters the core. When the NS enters the core it launches jets that explode the star and expand almost freely along the polar directions that contain a small amount of mass. This explains the fast rise to maximum and the fast ejecta observed at early times. The slower later time ejecta is the more massive equatorial outflow.

We showed that in many cases the NS can spiral-in all the way to core (Figs. \ref{fig:MassDensity8M}--\ref{fig:MassDensity15M}). The outcome depends on the stage of engulfment and on the poorly determined efficiency of the jets, that the NS might launch, in removing the common envelope. In the present study we qualitatively mimic the effect of jets by taking the common envelope $\alpha$ parameter to be $\alpha=3$. Namely, we took a case where the jets deposit to the common envelope twice as much energy as the orbital energy of the NS-core binary system. When we include the energy that we expect the jets to deposit to the envelope we find that there is a large parameter space for the formation of close NS binaries after the core explodes, but here we were interested in the cases when the NS enters the core.

The NS destroys the core, after entering it, by tidal torques and jets. We assume that in the final phase the core material forms an accretion disc around the NS. The disc launches jets that explode the star. We roughly estimated the accretion phase onto the NS during the explosion phase to last for a time of $\approx 10^{3} \s$, during which the average mass accretion rate is $\approx 10^{-4} \rmModot \s^{-1}$ (section \ref{sec:accretion}). In Fig. \ref{fig:Accretion} we show that the interaction can supply this accretion rate.
 
We note that even that the NS enters and destroys a helium core, in the centre of the ejecta there might be a carbon/oxygen rich zone resulting from two sources. The first one is the burning of helium that might have started already. This can supply at most about ${\rm several} \times 0.01 M_\odot$. The second one is the nucleosynthesis of carbon and oxygen in the post-shock zones of the shocks that the jets excite in the dense parts of the helium core. We will determine the nucleosynthesis in future 3D hydrodynamical simulations.

In general, the properties of the giant star when it swallows the NS determine the characteristics of a resultant CEJSN or CEJSN impostor. There can be several qualitative different outcomes, some of which we list in Table \ref{Table:Scenarios}. Only full 3D hydrodynamical simulations with a very high resolution around the NS, as to allow for the formation of an accretion disc, can determine the outcome of the interaction, and to find whether the speculative polar CEJSN scenario we proposed here for AT2018cow can take place. 
 
\ifmnras
\section*{Acknowledgments}
\else
\vspace{0.8cm}
\fi

We thank an anonymous referee for useful comments. 
This research was supported by the Asher Fund for Space Research at the Technion, and the Israel Science Foundation. A. Grichener was supported by The Rothschild Scholars Program - Technion Program for Excellence. A. Gilkis gratefully acknowledges the support of the Blavatnik Family Foundation.

\ifmnras
\bibliographystyle{mnras}
\fi

\label{lastpage}
\end{document}